# Experiences of Engineering Grid-Based Medical Software

F. Estrella, T. Hauer, R. McClatchey, M. Odeh, D Rogulin & T. Solomonides

Centre for Complex Cooperative Systems, CEMS Faculty, Univ. of the West of England, Coldharbour Lane, Frenchay, Bristol BS16 1QY, United Kingdom
Telephone: +44 117 328 3761, FAX: +44 117 344 3155
Email: {Richard.McClatchey, Mohammed.Odeh, Tony.Solomonides}@uwe.ac.uk,

{Florida.Estrella, Tamas.Hauer, Dmitry.Rogulin}@cern.ch

**Summary**

*Objectives:* Grid-based technologies are emerging as potential solutions for managing and collaborating distributed resources in the biomedical domain. Few examples exist, however, of successful implementations of Grid-enabled medical systems and even fewer have been deployed for evaluation in practice. The objective of this paper is to evaluate the use in clinical practice of a Grid-based imaging prototype and to establish directions for engineering future medical Grid developments and their subsequent deployment.
*Method:* The MammoGrid project has deployed a prototype system for clinicians using the Grid as its information infrastructure. To assist in the specification of the system requirements (and for the first time in healthgrid applications), use-case modelling has been carried out in close collaboration with clinicians and radiologists who had no prior experience of this modelling technique. A critical qualitative and, where possible, quantitative analysis of the MammoGrid prototype is presented leading to a set of recommendations from the delivery of the first deployed Grid-based medical imaging application.
*Results:* We report critically on the application of software engineering techniques in the specification and implementation of the MammoGrid project and show that use-case modelling is a suitable vehicle for representing medical requirements and for communicating effectively with the clinical community. This paper also discusses the practical advantages and limitations of applying the Grid to real-life clinical applications and presents the consequent lessons learned.
*Conclusions:* The work presented in this paper demonstrates that given suitable commitment from collaborating radiologists it is practical to deploy in practice medical imaging analysis applications using the Grid but that standardisation in and stability of the Grid software is a necessary pre-requisite for successful healthgrids. The MammoGrid prototype has therefore paved the way for further advanced Grid-based deployments in the medical and biomedical domains.

## 1. Introduction

The past decade has witnessed order of magnitude increases in computing power and data storage capacity, giving birth to new applications which can handle large data volumes of increased complexity. Similar increases in network speed and availability pave the way for applications distributed over the web, carrying the potential for better resource utilization and on-demand resource sharing. Medical informatics is one of the areas in which these technologically revolutionary advances could bring significant benefit both for scientists' research study and clinicians' everyday work. Recently there has been much excitement in the parallel systems community as well as that of distributed database applications in the emergence of 'The Grid' as a promising platform for scientific and medical collaborative computing.

The Grid is a new paradigm for distributed computing defined as the "flexible, secure, coordinated resource sharing among dynamic collections of individuals, institutions and resources" [1]. Geographically separated yet working together to solve a problem, groups of people can harness the collection of resources provided by the participants and use the shared environment of the Grid within the boundaries of so-called Virtual Organizations (VO).
In essence the Grid:
- provides a virtual platform for large-scale, resource-intensive, and distributed applications;
- offers a connectivity environment allowing management and coordination of diverse and dispersed resources;
- enables access to increased storage capacity and computing power;
- provides mechanisms for sharing and transferring large amounts of data as well as aggregating distributed resources for running computationally expensive procedures; and
- utilizes a common infrastructure based on open standards thus providing a platform for interoperability and interfacing between different Grid-based applications from the particular domain.

Grid computing holds the promise of harnessing extensive computing resources located at geographically dispersed locations which can be used by a dynamically configured group of collaborating institutions; consequently, it defines a suitable platform on which to base distributed medical informatics applications. In particular the Grid can address some of the following issues relevant to medical domains:

*Data distribution*: The Grid provides a connectivity environment for medical data distributed over different sites. It solves the location transparency issue by providing mechanisms which permit seamless access to and the management of distributed data. These mechanisms include services which deal with virtualization of distributed data regardless of their location.

*Heterogeneity:* The Grid addresses the issue of heterogeneity by developing common interfaces for access and integration of diverse data sources. Such generic interfaces for consistent access to existing, autonomously managed databases that are independent of underlying data models are defined by the Global Grid Forum Database Access and Integration Services (GGF-DAIS) [2] working group. These interfaces can be used to represent an abstract view of data sources which can permit homogeneous access to heterogeneous medical data sets.

*Data processing and analysis*: The Grid offers a platform for transparent resource management in medical analysis. This allows the virtualization and sharing of all resources (e.g. computing resources, data storage, etc.) connected to the grid. For handling computationally intensive procedures (e.g. CADe), the platform provides automatic resource allocation and scheduling and algorithm execution, depending on the availability, capacity and location of resources.

*Security and confidentiality*: Enabling secure data exchange between hospitals distributed across networks is one of the major concerns of medical applications. Grid addresses security issues by providing a common infrastructure for secure access and communication between grid-connected sites. This infrastructure includes authentication and authorization mechanisms, among other things, supporting security across organizational boundaries.

*Standardization and compliance*: Grid technologies are increasingly being based on a common set of open standards (such as XML, SOAP, WSDL, HTTP etc.) and this is promising for future medical image analysis standards.

In other words, Grid computing has the capacity to resolve many of the exceptional difficulties encountered in medical informatics by allowing medical doctors and researchers to collaborate without having to co-locate, thereby providing transparent access to data and computing resources. To date, however, there have been few projects that have attempted to deliver Grid computing to clinicians and there is no practice-based evidence or guidelines as to where and how the Grid can benefit clinicians. Furthermore there are few articles in the established medical informatics journals that have covered aspects of Grid computing [3], [4].

In this paper, we report on how a Grid architecture has been used to provide both computing power and a distribution platform to a community of radiologists spanning multiple medical institutions, even across borders, and in so doing to investigate the particular issues surrounding the implementation and use of Grid technologies in a clinical environment. MammoGrid exploited existing and emerging technologies to build a large-scale database of mammograms and associated metadata that can be used to investigate healthcare applications and to explore the potential of the Grid to support effective co-working between healthcare professionals [5, 6]. The use of digitized radiological images (mammograms) enabled linkage of distant centres for the first time in a "radiological virtual organization". The MammoGrid project aimed to demonstrate that through such a virtual organization a Grid infrastructure can support collaborative medical image analysis, and to enable radiologists to share standardized mammograms, compare diagnoses (with and without computer-aided detection), and perform epidemiological research studies across national boundaries. This paper highlights the specific constraints apparent in Grid-based distributed radiology and outlines how established software engineering techniques can be married with emerging Grid technologies to provide a first Grid-based mammogram analysis system.

The paper is structured as follows. First, we present the rationale behind the MammoGrid project and we restate its aims and objectives to provide justification for the approach followed in its development. In the next section, we investigate how this approach was constrained by two major factors: the nature of the clinical domain and the rapidly changing Grid research environment. Then, we describe (in section 3) how we approached the MammoGrid research and development process to address these constraints. Emphasis on the requirements engineering phase of the project and the construction of a set of clinical use-cases for the capture of the radiologists' system needs are presented in section 4. Next, in section 5, we identify the outcomes of the design and development phases of the MammoGrid project and outline its prototyping strategy. Then in section 6 the service-oriented architecture of the delivered prototype is presented and discussed along with its suitability through a set of clinical tests. This is then followed by section 7, where the main lessons from undertaking a software engineering approach in the MammoGrid project are summarized. Finally, we draw conclusions (in section 8) on the software engineering process and give guidelines for future development and deployment of Grid-based systems in the medical domain.

## 2. Background

The Fifth Framework EU-funded MammoGrid project [6] aimed to apply the Grid concept to mammography, including services for the standardization of mammograms, computer-aided detection (CADe) of salient features, especially masses and 'microcalcifications', quality control of imaging, and epidemiological research including broader aspects of patient data. In doing so, it attempted to create a paradigm for practical, Grid-based healthcare-oriented projects, particularly those which rely on imaging. There are, however, a number of factors that make patient management based on medical images particularly challenging. Often very large quantities of data, with complex structures, are involved (such as 3-D images, time sequences, multiple imaging protocols etc.). Also, clinicians rarely analyse single images in isolation but rather in the context of metadata[1]. Metadata that may be required are clinically relevant factors such as patient age, exogenous hormone exposure, family and clinical history; for the population, natural anatomical and physiological variations; and for the technology, image acquisition parameters, including breast compression and exposure data. Thus any database of images developed at a single site may not contain enough exemplars in response to any given query to be statistically significant. Overcoming this problem implies constructing a very large and federated database, which can transcend national boundaries. However this will necessitate specialist image processing algorithms – for example, computationally heavy tasks operating on large files of images – which in turn place significant requirements on storage space, CPU power and/or network bandwidth on all participating hospitals, unless appropriate sharing of computing resources is arranged. Realising such a geographically distributed (pan-European) database therefore necessitates a Grid infrastructure, and the construction of a prototype model which would push emerging Grid technology to its limits.

The MammoGrid project was carried out between mid 2002 and the end of 2005 and involved hospitals and medical imaging experts and academics in the UK, Italy and Switzerland with experience of implementing Grid-based database solutions. A key deliverable of the project was a prototype software infrastructure based on an open-source Grid 'middleware' (i.e. software that enables an underlying Grid infrastructure to host domain applications) and a service-oriented database management system that is capable of managing federated mammogram databases distributed across Europe. The proposed solution was a medical information infrastructure delivered on a service-based, grid-aware framework, encompassing geographical regions of varying clinical protocols and diagnostic procedures, as well as lifestyles and dietary patterns. The prototype will allow, among other things, mammogram data mining, diverse and complex epidemiological studies, statistical and (CADe) analyses, and the deployment of versions of the image standardization software. It was the intention of MammoGrid to get rapid feedback from a real clinical community about the use of such a simple Grid platform to inform the next generation of Grid projects in healthcare.

The clinical workpackages encompassed in MammoGrid prototypes address three selected clinical problems:
- Quality control: the effect of image variability, due to differences in acquisition parameters and processing algorithms, on clinical mammography;
- Epidemiological studies: the effects of population variability, regional differences such as diet or body habitus and the relationship to mammographic density (a biomarker of breast cancer) which may be affected by such factors;
- Support for radiologists, in the form of tele-collaboration, second opinion, training and quality control of images.

Other initiatives against which MammoGrid may be compared include: the eDiamond [7] project in the UK, and the NDMA [8] project in the US. The MammoGrid approach shares many similarities with these projects, but in the case of the NDMA project (one of whose principal aims is to encourage the adoption of digital mammography in the USA) its database is implemented in IBM's DB2 on a single server. The MammoGrid project federates multiple (potentially heterogeneous) databases as its data store(s). The Italian INFN project GP-CALMA (Grid Project CALMA) [9] has focused on a Grid implementation of tumour detection algorithms to provide clinicians with a working mammogram examination tool. MammoGrid uses aspects of the CALMA project in its computer-aided detection of microcalcifications.

---

[1] 'Metadata' is used inclusively to encompass associated data (such as patient information), summary data (such as breast density) and metadata proper, such as information about the logical or physical distribution of the data.

More recent Grid-based research includes the BIRN [10] project in the US, which is enabling large-scale collaborations in biomedical science by utilizing the capabilities of emerging Grid technologies. BIRN provides federated medical data, which enables a software 'fabric' for seamless and secure federation of data across the network and facilitates the collaborative use of domain tools and flexible processing/analysis frameworks for the study of Alzheimer's disease. The INFOGENMED initiative [11] has given the lead to projects in moving from genomic information to individualized healthcare using data distributed across Europe. Finally the CDSS [12] project is a system which uses knowledge extracted from clinical practice to provide a classification of patients' illnesses, implemented on a Grid platform.

From the outset, the MammoGrid project posted its objectives in terms of the promised radiological and epidemiological applications, but not in terms of new Grid technology. Its technology attitude has largely been one of re-use, not invention or development; only where required functionality was missing was there a need to implement new Grid services. An information infrastructure to integrate multiple mammogram databases is clearly needed to enable clinicians to develop new common, collaborative and co-operative approaches to the analysis of mammographic images as is evident by the clinical evaluation that took place towards the end of the MammoGrid project.

## 3. The Development Environment

The development and deployment of the MammoGrid prototypes was carried out between 2002 and 2005 by a group of software engineering researchers from the University of the West of England (UWE, UK) and from European Centre for Particle Physics (CERN, Switzerland), groups of medical imaging experts from Mirada Solutions (UK) and the Universities of Oxford, Pisa and Sassari, and research radiologists from the University hospitals of Udine (Italy) and Addenbrookes, Cambridge (UK). The approach followed was very pragmatic in nature rather than one which rigorously followed a traditional software engineering process and was loosely based around an evolutionary prototyping philosophy; nevertheless, in its early stages, the project conducted its requirements capture by utilising aspects of the Rational Unified Process Model (RUP) [13] and by establishing a set of use-cases which subsequently were used as part of the clinical evaluation of the project's outcomes.

Due to the nature of the development environment there were a number of constraints on the software engineering process. *First*, the development was research-oriented both in terms of the maturity of the still-emerging Grids middleware and the novelty of the MammoGrid services for the medical community. This constraint led to several challenges: the need to raise the clinicians' awareness of Grid technologies and the use of requirements modelling techniques while managing their expectations of what these technologies might deliver; the necessity to cater for frequent releases of new underlying software technologies while at the same time adhering to existing medical standards and protocols, and the need to re-train researchers 'on-the-job' as new middleware became available.

*Second*, the MammoGrid project was carried out under both tight manpower and time restrictions. This necessitated careful project management of the collaboration between busy clinicians, software engineers, developers and computer science doctoral students. As a consequence, this required significant participation from the user community (radiologists, radiographers) especially during the requirements engineering phase of the project along with frequent feedback and validation of project findings (mainly through systems 'walkthroughs' or sometimes 'stomp-overs') with the software engineers and researchers. *Third*, the project required research and development effort spread over eight institutes located in the UK, Italy and Switzerland, a substantial challenge to project management. It was therefore necessary to delineate clear task boundaries and to establish inter-task dependencies, so that explicit responsibilities for the production of deliverables were established and that, as a group, the project respected those responsibilities and adhered to delivery milestones. During the early stages of the project, strong bonds of trust and mutual respect were built between project members. Also, agreed project management structures and deliverables/milestones were put in place. Project members remained committed to the common goals despite shortages of resources and despite differences in priorities between research and commercial partners.

*Finally*, during the later stages of the project the deployment and testing of the system were subject to delivery constraints of software from commercial partners and subject to the maturity and stability of the underlying Grid technologies as it evolved during the project. This inevitably led to delays in the successful integration of the final MammoGrid prototype and the testing and validation of the clinical use-cases.

Constraints on the requirements elicitation, specification and validation phases included the limited time available with domain experts, the geographic distance between the various stakeholders and hence the episodic nature of meetings. In the course of a visit of several days, domain experts could make themselves available for relatively frequent but rather brief meetings. Domain experts had no previous exposure to the kind of model used in software development. Software engineers on the project had some appreciation but very little experience to the particular problems of mammography and breast cancer screening prior to this exercise. Moreover, the requirements team had to span the space between radiologists, Grid experts and medical image processing specialists, whether those working on the specification of the local workstation or on the CADe software.

## 4. The Requirements Engineering Phase.

The MammoGrid project was driven by the requirements of its user community – Udine and Cambridge hospitals – along with medical imaging expertise from Oxford. The ultimate objective of the requirements engineering phase was to obtain an agreed, validated and essentially stable requirements specification document for the project. Two core objectives for the project followed immediately from its scope and definition:

- The support of clinical research studies through access to and execution of algorithms on physically large, geographically distributed and potentially heterogeneous sets of (files of) mammographic images, just as if these images were locally resident; and

- The controlled and assured access of educational and commercial companies to distributed mammograms for testing novel medical imaging diagnostic technologies in scientifically acceptable clinical trials that fulfil the criteria of evidence-based medical research.

To facilitate requirements specification, a number of meetings took place between software engineers and radiologists at Udine and Cambridge to elicit, and then analyse and specify the functional requirements of the end-user radiologists and radiographers (radiology technicians) in addition to product-related non-functional requirements. Use-case and conceptual data (object-oriented) models were incrementally and iteratively developed and validated as the main requirements models followed by dynamic interaction and state transition diagrams. Parallel to the requirements elicitation activities, the hardware and software requirements were established. Meanwhile, the logical view of the application architecture was developed following iterations on activities in the requirements and design workflows of RUP. The requirements were specified by a group of UWE software engineers working with domain experts from Mirada Solutions and the participating hospitals.

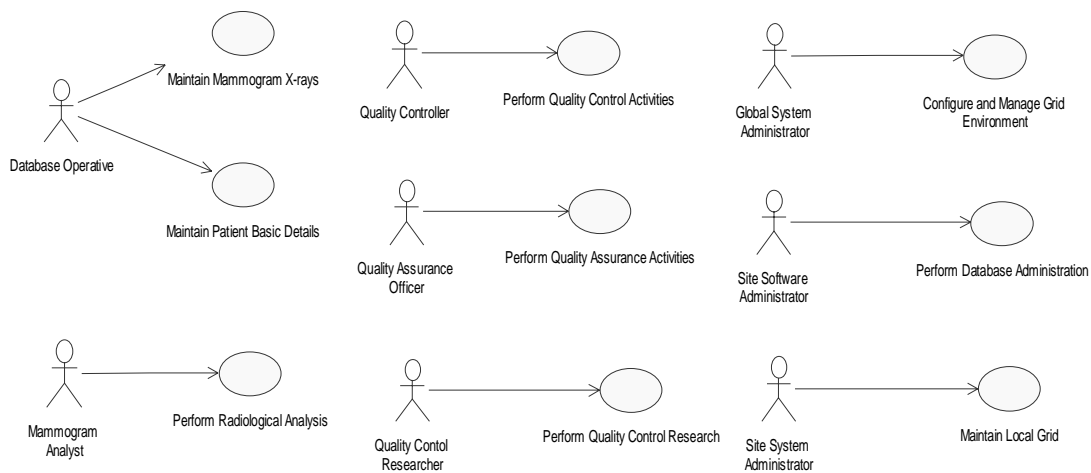

**Figure 1 – The MammoGrid system level use-case view.**

These activities resulted in identifying major use-case scenarios in the use of a distributed database of mammograms deployed across a pan-European grid that were validated with the domain specialists and then later used to prove the project prototype(s) in clinical evaluation. Problem domain entities (classes) were identified and described in addition to documenting relationships between such entities, resulting in a stable and validated conceptual class model which has since evolved as the logical class model. The system level use-case model of MammoGrid is shown in Figure 1. A detailed example, the essential function of mammogram analysis carried out by a "mammogram analyst" (one of key system actors, normally enacted by a clinician) is briefly modelled and presented in Figure 2 using the core use-case "Perform Radiological Analysis". Further details of the use-case model can be found in [14].

Use-case analysis and modelling identified the major actors of the system, differentiating these from job roles or individuals, and investigated how they impacted principal system functions across several scenarios. This was further iteratively validated in a process that involved all stakeholders, including users, researchers and developers; this strengthened the cohesion of the project, provided a common visual language for communication and problem representation and led to an agreed requirements specification.

The main requirements elicitation methods used were semi-structured interviews and strictly non-participant observation of medical procedures; with appropriate permissions, the team observed such procedures as basic mammography, ultrasound-guided biopsy, breast MRI, reading of mammograms and other images, and X-ray examination of biopsy specimens. The first resulting requirements models (i.e. the use-case models) were established and then iteratively and incrementally presented to radiologists at Udine and CADe experts from Pisa and Sassari (with contributions from a further private hospital in Torino), then to radiologists at Cambridge, then to imaging specialists and finally to two project plenary meetings. In parallel, the team from Mirada Solutions constructed the 'acquisition system' prototype and defined related data structures. It was then possible to cross refer and thus to validate the data requirements emerging from the use cases. The acquisition system remained an architectural component of the MammoGrid.

**Perform Radiological Analysis**

*Pre-Conditions:*
    User-Authentication
*Non-functional requirements:*

CADe Software Interface Requirements
*Flow of Events:*
    as per selection of the Mammogram Analyst to
    link to the appropriate extension point below
*Extension Points:*
    (1) View Mammogram and Patient Details
    (2) Annotate Mammograms and Patient Details
        (2.1) Diagnose Study
        (2.2) Diagnose Series
        (2.3) Annotate Image
        (2.4) Request CADe
        (2.5) Link Annotations
        (2.6) Request CADe in Mammogram Region
    (3) Execute Radiological Queries
        (3.1) Formulate Radiological Query
        (3.2) Refine Radiological Query
*Alternative Flows:*
    (1) Unsuccessful User Authentication
    (2) CADe Interface Error
    (3) Invalid Query Selections
*Post-Conditions:*
    (1) Mammogram Image Annotated
    (2) Patient Details Changed
    (3) Results of Query Execution (Grid)

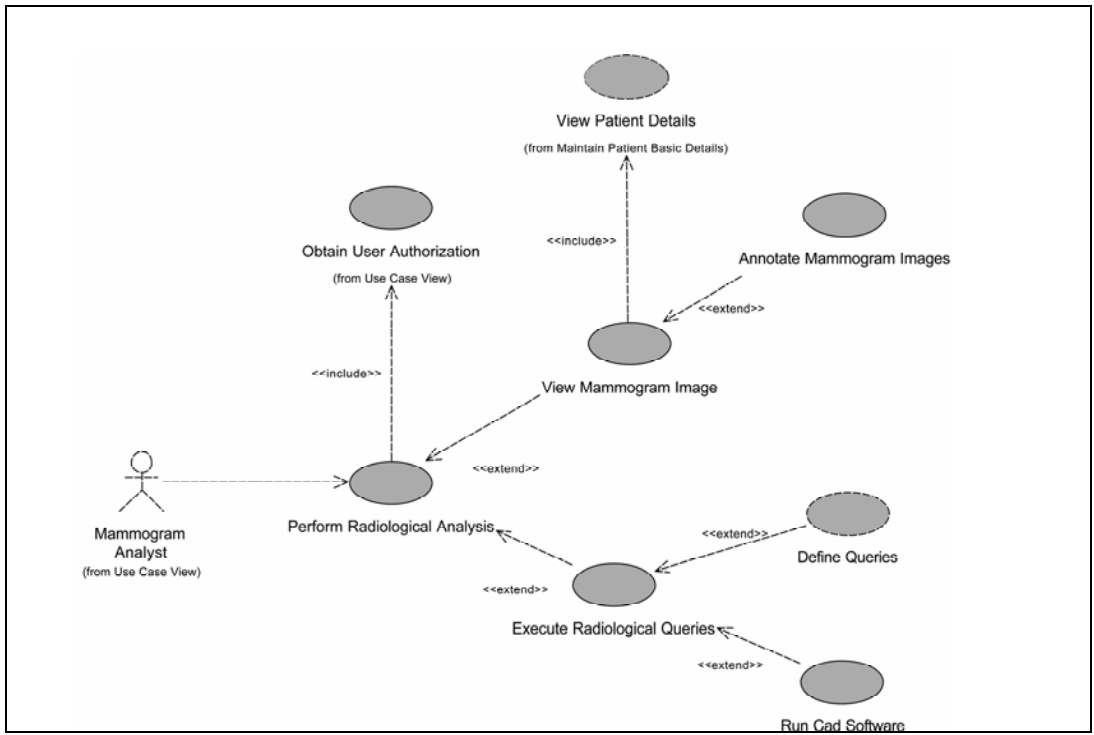

**Figure 2 – Use case hierarchy and diagram**

An analysis of project non-functional requirements (NFRs) was conducted including product-related, organization and process requirements, external constraints, such as confidentiality, and interface specifications. In this analysis, constraints on the process of mammogram study, such as usability, reliability, robustness and security were investigated and specified as a means to assess the degree of adherence to these requirements at implementation time. In addition, the impact of product-related NFRs on architecture selection and specification was investigated.

The MammoGrid project has certain non-standard characteristics such as:

- a wide diversity of backgrounds among problem domain specialists (radiologists, radiographers, epidemiologists, medical imaging experts);
- the application domain itself, roughly speaking, the construction of the evidence base for radiological practice in mammography;
- the geographically dispersed locations of the different parties involved in the project; and
- the need to establish *ab initio* the use of a modelling language, UML [15], (in particular, use-cases) with corresponding modelling, validation, and requirements management tools.

**5.     The Design and Deployment Phases.**

As set out in its goals, the MammoGrid project concentrated on applying existing Grid middleware rather than developing new Grid software: the design philosophy adopted in the project focused on services that address user requirements for collaborative mammogram analysis. One of the main deliverables of the project was an interface between the radiologist's image analysis workstation and the 'MammoGrid Information Infrastructure' (MII) based on the philosophy of a Grid. This enabled radiologists to query images across a widely distributed federated database of mammographic images and to perform epidemiological and CADe analyses on the sets of returned images. In delivering the MII, the MammoGrid project has customised and, where necessary, enhanced and complemented Grid software for the creation of a medical analysis platform. The approach that is being followed in the project is therefore twofold: to provide an MII based on a service-oriented architecture [16], [17] and a metadata and query handler coupled to a 'front-end' to ensure that both patient data and images remain appropriately associated and that metadata based searches are effectively handled. The MII has been fully specified and a prototype delivered, in which a set of medical imaging services is implemented to manage the federation of distributed mammograms [18].

To encourage re-use the MammoGrid software was delivered through a set of evolving prototypes following a form of 'spiral model' [19] development (including 'stages' of planning, specification, evaluation, and development for each prototype version) in which the clinical user community provided input in each loop of the spiral. Release of the staged prototypes was planned to coincide with project milestones and the delivery of tested MammoGrid services. Involvement of the clinicians helped to maintain their engagement with the project at a stage when they could not yet draw benefit from any tangible system. This ensured commitment to project deliverables and enabled the software developers to gain a deeper understanding of the actual system requirements of the clinicians; these were important benefits of this design and its implementation approach. This strategy also enabled the project to cope easier with the multiple versions of the underlying Grid software that emerged during the lifetime of the project as well as with regular updates to the clinical workstation provided by Mirada Solutions.

The MammoGrid project has recently delivered its final proof-of-concept prototype enabling clinicians to store digitized mammograms along with appropriately anonymized patient metadata; the prototype provides controlled access to mammograms both locally and remotely stored. A typical database comprising several thousand mammograms has been created for user tests of clinicians' queries. The prototype comprises

- a high-quality clinician visualization workstation (used for data acquisition and inspection);
- an imaging standard-compliant interface to a set of medical services (annotation, security, image analysis, data storage and querying services) residing on a so-called 'Grid-box'; and
- secure access to a network of other Grid-boxes connected through Grids middleware.

## 6. Clinical Evaluation of the MammoGrid Prototype.

The evaluation of the MammoGrid prototype was conducted in the final months of the project and was driven by assessing the achievement of the overall project objectives. The evaluation process therefore concentrated on the following aspects:
- The establishment and deployment of a MammoGrid virtual organisation;
- The evaluation of the service oriented approach with the emphasis on clinical services provided by the Grid middleware layer;
- The use of the clinical workpackages (by senior radiologists) specified before the start of the project to drive the qualitative and quantitative evaluation of the implemented use-cases in order to provide
- Feedback for future 'healthgrid' projects on the clinical collaborative nature of the adopted approach in the form of lessons learnt.

The discussion below summarises the main outcomes of evaluating the MammoGrid's prototype in light of the above criteria.

To allow for the evaluation of the final prototype at the test clinical sites, which was the first attempt at studying the use of a Grid-based cross-national database by practicing radiologists, a MammoGrid Virtual Organization (MGVO) was established and deployed (as shown in figure 3). The MGVO was composed of three mammography centres – Addenbrookes Hospital, Udine Hospital, and Oxford University. These centres were autonomous and independent of each other with respect to their local data management and ownership. The Addenbrookes and Udine hospitals had locally managed databases of mammograms, with several thousand cases between them (see table 1 below). As part of the MGVO, registered clinicians had access to (suitably anonymized) mammograms, results, diagnosis and imaging software from the other centres. Access was coordinated by the MGVO's central node at CERN.

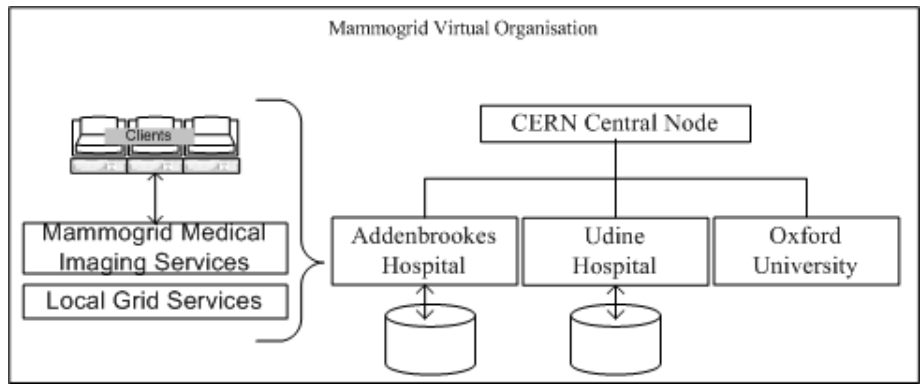

**Figure 3 – The MammoGrid Virtual Organisation (MGVO)**

In order to minimise development and maximise re-use of existing Grid software, the adopted middleware solution was the ALICE Environment (AliEn) [20] component of the EGEE-gLite middleware [21] i.e. the grid middleware of the EU-funded EGEE project [22]. The service-oriented approach adopted in MammoGrid permitted the interconnection of communicating entities, called services, which provided capabilities through exchange of messages. The services were 'orchestrated' in terms of service interactions: how services were discovered, how they were invoked, what could be invoked, the sequence of service invocations, and who could execute them.

The MammoGrid Services (MGS) are a set of services for managing mammography images and associated patient data on the grid. Figure 4 illustrates the services that made up the MGVO. (For simplicity, Oxford University has not been included).

The MGS are: (a) *Add* for uploading files (DICOM [23] images and structured reports) to the MGVO; (b) *Retrieve* for downloading files from the grid system; (c) *Query* for querying the federated database of mammograms; (d) *AddAlgorithm* for uploading executable code to the Grid; (e) *ExecuteAlgorithm* for executing grid-resident executable code on grid-resident files on the Grid system; and (f) *Authenticate* for logging into the MGVO. See [18] for further details.

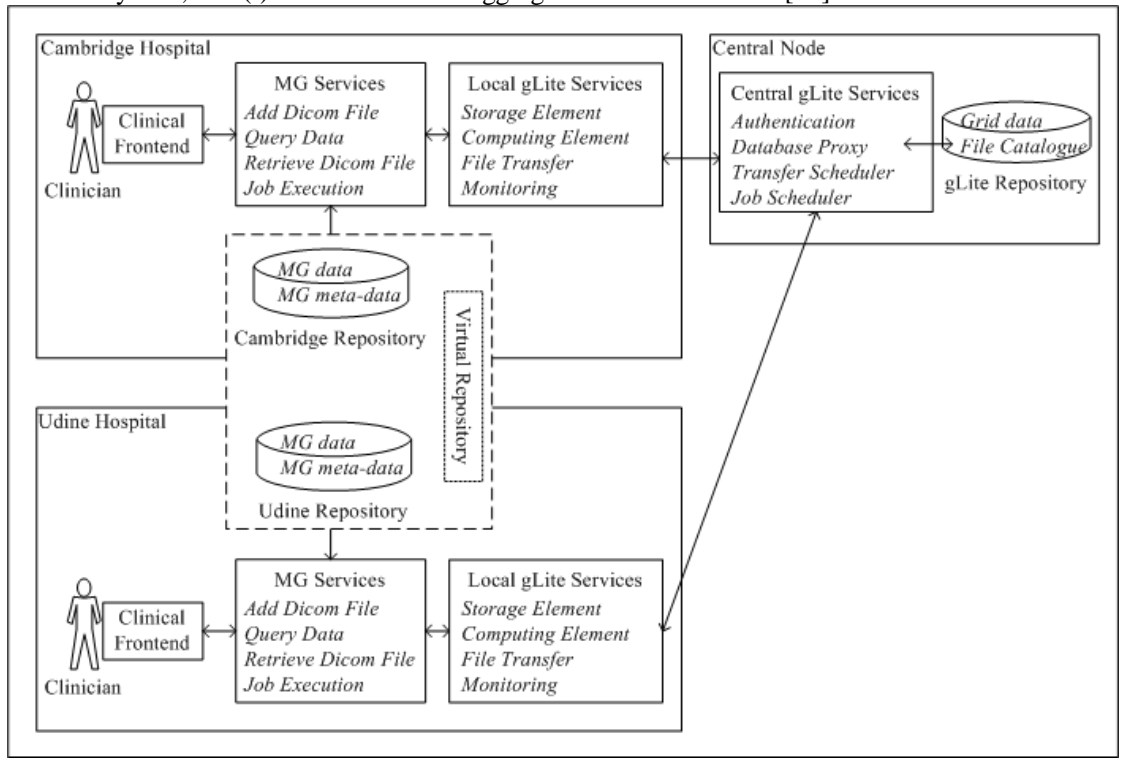

**Figure 4 – The MammoGrid Services in the MGVO.**

Evaluation of the MammoGrid prototype took place using the MGVO over a set of clinical workpackages performed by senior radiologists at Addenbrookes and Udine hospitals. The MammoGrid Virtual Organisation encompassed data accessible to the radiologists at the hospitals, as well as at Oxford University and CERN. The evaluation comprised the qualitative and, where possible, quantitative assessment of the use-cases captured during the requirements elicitation phase of the project (detailed in section 4). The domain of the evaluation reflected on the key elements of the clinical workpackages, as identified in section 2, and these are::
- *Quality control:* the effect of image variability, due to differences in acquisition parameters and processing algorithms, on clinical mammography;
- *Epidemiological studies:* the effects of population variability, regional differences such as diet or body habitus and the relationship to mammographic density (a biomarker of breast cancer) which may be affected by such factors;
- *Support for radiologists*, in the form of tele-collaboration, second opinion, training and quality control of images.

During this clinical evaluation, the radiologists were able to view raw image data from each others' hospitals and were able to second-read Grid-resident mammograms and to separately annotate the images for combined diagnosis. This demonstrated the viability of distributed image analysis using the Grid and showed considerable promise for future health-based Grid applications. Despite the anticipated performance limitations that existing Grid software imposed on the system usage, the clinicians were able to discover new ways to collaborate using the virtual organisation. These included the ability to perform queries over a virtual repository spanning data held in Addenbrookes and Udine hospitals and joint analyses thereof.

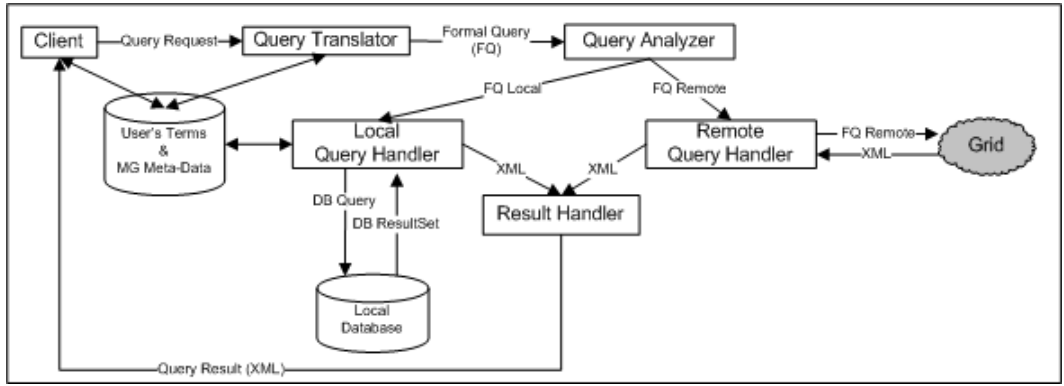

**Figure 5 – Clinical Query Handing in MammoGrid.**

Following the 'Perform Radiological Analysis' use-case scenario shown in figure 2, clinicians defined their mammogram analysis in terms of queries they wished to be resolved across the collection of data repositories. Queries were categorized into simple queries (mainly against associated data stored in the database as simple attributes) and complex queries which required *derived data* to be interrogated or *special purpose algorithms* (e.g. for detection of abnormalities) to be executed on a (sub-)set of distributed images. One important result was that image and data distribution were transparent for radiologists, and hence complex queries were formulated and executed as if the associated data and images were locally resident. Queries were executed at the location where the relevant data resided, i.e. sub-queries were moved to the data, rather than large quantities of data being moved to the clinician, which could have been prohibitively expensive given the volume of the data involved. Figure 5 illustrates how queries were handled in MammoGrid.

The Query Analyzer took a formal query representation and decomposed it into (a) a formal query for local processing, and (b) a formal query for remote processing. It then forwarded these decomposed queries to the Local Query Handler and the appropriate Remote Query Handler for the resolution of the request. The Local Query Handler generated the corresponding query language statements (e.g. SQL) in the query language of the associated Local DB. The result set was converted to XML and then routed to the Result Handler. The Remote Query Handler is a portal for propagating queries and their execution results between sites. This handler forwarded the formal query for remote processing to the Query Analyzer of the remote site. The remote query result set was converted to XML and routed to the Result Handler. For detail see also [18]. At the time of writing this paper, the database is continuing to grow and currently holds:

| Site | Number of Patients | Number of Image Files | Number of SMF Files | Associated Database Size | File Storage Size |
|---|---|---|---|---|---|
| Cambridge | 1423 | 9716 | 4815 | 14.0 Mb | 260 Gb |
| Udine | 1479 | 17285 | 8634 | 23.5 Mb | 220 Gb |
| **TOTAL:** | **2902** | **27001** | **13449** | **37.5 Mb** | **480 Gb** |

**Table 1 : Virtual Repository size of the MammoGrid prototype.**

The average processing time for the core services was: (1) Add a 8Mb DICOM file approximately 7 seconds (2) Retrieve a 8Mb DICOM file from a remote site approximately 14 seconds (3) SMF workflow of ExecuteAlgorithm and Add around 200 seconds. The evaluation, carried out in mid 2005 on a *subset* of the currently available data revealed that for querying:

| Query | Cambridge | Udine | Num images | Num patients |
|---|---|---|---|---|
| By Id: Cambridge patient | 2.654 sec | 2.563 sec | 8 | 1 |
| By Id: Udine patient | 2.844 | 3.225 | 16 | 1 |
| All female | 103 | 91 | 12571 | 1510 |
| Age [50,60] and ImageLaterality=L | 19.489 | 22.673 | 1764 | 357 |

**Table 2 : Data query performance of the MammoGrid prototype.**

As a direct result of their satisfaction with the MammoGrid evaluation, clinicians continue in the process of scanning and annotating cases that contribute to several ongoing medical studies. These include (1) Cancers versus Control study: breast density study using SMF standard, (2) Dose/Density study: exploring the relationship between mammographic density, age, breast size and radiation dose, and (3) CADe and validation of SMF in association with CADe. These studies continue to show how health professionals can work together without co-locating. And, most importantly the collaborative approach pursued in MammoGrid has already identified new ways in which clinicians can work together using a common Grid-based repository which were hitherto not possible. For example, the use of the SMF [24] algorithm on data supported by MammoGrid and accessible to radiologists in Cambridge and Udine for the purposes of joint mammogram analysis has directly led to results being recently submitted to the European Radiology Journal [25].

In summary, during the final months of the project the clinicians have evaluated the MammoGrid prototype across two applications. *First*, the project has facilitated the use of the SMF software to measure breast density. The clinical project, designed jointly by Cambridge and Udine, explored the relationship between mammographic density, age, breast size, and radiation dose. In this project, breast density has been measured by SMF and compared with standard methods of visual assessment. Heights, weights, and mass indicators are used in an international comparison, but a richer dataset would be needed to study effects of lifestyle factors such as diet or HRT (Hormone Replacement Therapy) use between the two national populations. *Second*, the University of Udine led a project to validate the use of SMF in association with CADe from the CALMA project [9]. Cancers and benign lesions have been supplied from the clinical services of Udine and Cambridge to provide the benchmarking and the set of test cases. Cancer cases include women whose unaffected breast will serve the density study to provide cases for the CADe analysis from the affected side mammogram. MammoGrid has demonstrated that these new forms of clinical collaboration can be supported using the Grid [26].

*Furthermore* a strong collaboration has been established through the evaluation phase of the project between radiologists active in breast cancer research and academic computer scientists with expertise in the applications of grid computing. The success of the evaluation has led to interest from outside companies and hospitals, with one Spanish company, Maat GKnowledge [27], looking to deploy a commercial variant of the system in three hospitals of the Extremadura region in Spain. Maat GKnowledge aims to provide the Extremadura doctors with the ability to verify test results, to obtain second opinions and to make use of the clinical experience acquired by the hospitals involved in the MammoGrid project. They then aim to scale the system up and to expand it to other areas of Spain and then Europe. With the inclusion of new hospitals, it is proposed that the database will increase in coverage with clinical knowledge increasing in relevance and accuracy, and thus enabling larger and more refined epidemiological studies. Consequently, clinicians will be provided with a significant data set to serve better their investigations in the domain of cancer prevention, prediction and diagnoses. This is expected to result in improved research quality as well as improved citizen access to the latest healthcare technologies. Further details of the clinical evaluation of MammoGrid and its exploitation plans can be found in [28].

## 7. Lessons Learned

The nature of the project and its particular constraints of multi-disciplinarity, dispersed geographical development, large discrepancies in participants' domain knowledge (whether of software engineering techniques or of breast cancer screening practice), and the novelty of the Grid environment provide experiences from which other Grid-based medical informatics projects can benefit. We summarise below some of the main lessons that can be learned in this context.

*First*, the project was particularly fortunate with selection of its medical partners. In general, the medical environment is very risk-averse, conservative in nature and reluctant to adopt new technologies without significant evidence of tangible benefit. It is therefore important in Grid system prototyping to identify a suitable user community in which new technologies (such as Grid –resident medical databases) can be evaluated. In the case of MammoGrid we have had real commitment from the radiology community in the project's requirements definition and analysis, implementation and evaluation and this was crucial to the success of the project. The data samples used were of a sensitive nature and required both ethical clearance from participating institutions and anonymization of the data and even then strictly for only research use in the project. Realistically many ethical obstacles remain to be tackled before clinicians can share sensitive patient data between institutes, never mind across national boundaries.

*Second*, it has become clear from our experiences that Grid middleware technology itself is still evolving, and this suggests that there is a clear need for standardization to enable production-quality systems to be developed. Despite the availability of toolsets such as the Globus 4.0 [29] the development of applications that harness the power of the Grid, as yet, requires specialist skills and is thus costly in terms of manpower. Only with the arrival of stable middleware and packaged Grid services will the development of medical applications become viable.

*Third*, the performance of existing middleware is also somewhat limited; the MammoGrid project had therefore to circumvent some of the delivered Grid services to ensure adequate system performance for its prototype evaluation. For example, the database of medical images was taken out of the Grid software to provide adequate response for MammoGrid query handling. The EGEE [22] project is addressing these technological deficiencies and improved performance of the middleware should consequently be delivered in the coming years.

*Fourth*, Grid technology for medical informatics is still in its infancy and needs some proven examples of its applicability; MammoGrid is the first such exemplar in practice. Equally, awareness of Grid technology and its potential (and current limitations) must still be raised in the target user communities such as Health, Biomedicine, and more generally life sciences.

*Fifth*, the project has indicated that it is possible to use modelling techniques (such as use-cases from UML) in a widely distributed, multi-disciplinary software engineering problem domain, provided a very pragmatic approach is used, where adopting a certain modelling technique is, to some extent, independent from the software development life cycle model being applied. The MammoGrid project has benefited significantly in its coordination, communication and commitment by utilizing the use-case model as the *lingua franca* during user requirement analysis and system design rather than following the disciplines of RUP to the letter.

*Sixth*, the evolutionary approach to system development work packages has mitigated the effects of the project constraints of a highly dynamic research-oriented environment in which novices and specialists in software engineering have worked together even though they may have been geographically separated.

Further areas that might promote the use of rigorous software engineering disciplines in the design of Grid-based software services are that of model-driven engineering [30] and the use of architecture descriptions [31] as the basis for the generation of Grid-wide services. These aspects are, however, outside the scope of the current project.

## 8. Future Directions and Conclusions

The MammoGrid Virtual Oragnisation (MGVO) is a distributed computing environment for harnessing the use of and access to massive amounts of mammography data across Europe. The MammoGrid approach used grid technologies, service-orientation, and database management techniques to federate distributed mammography databases allowing healthcare professionals to collaborate transparently without co-locating.

Furthermore, the MammoGrid project has delivered a Grid-enabled infrastructure which federates multiple mammogram databases across institutes. This permits clinicians to develop new common, collaborative and cooperative approaches to the analysis of mammography data. Using the MammoGrid they have been be able to quickly harness the use of massive amounts of medical image data to perform epidemiological studies, advanced image processing, radiographic education and ultimately, tele-diagnosis over communities of medical 'virtual organizations'. This was achieved through the use of Grid-aware services for managing (versions of) massively distributed files of mammograms, for handling the distributed execution of mammograms analysis software, for the development of imaging algorithms and for the sharing of resources between multiple collaborating medical centres.

In addition, the MammoGrid project has attracted attention as a paradigm for Grid-based imaging applications. While it has not solved all problems, the project has established an approach and a prototype platform for sharing medical data, especially images, across a Grid. In loose collaboration with a number of other European medical Grid projects (e.g. [7], [11], [12], [32], [33]), it is addressing the issues of informed consent and ethical approval, data protection, compliance with institutional, national and European regulations, and security [34].

In our view, the MammoGrid project paves the way for further research and development projects to meet the aims of the HealthGrid association [35] in the following respects:

- The identification of potential business models for medical Grid applications.
- Feedback to the Grid development community on the requirements of the pilot applications deployed by the European projects.
- Development of a systematic picture of the broad and specific requirements of physicians and other health workers when interacting with Grid applications.
- Dialogue with clinicians and those involved in medical research and Grid development to determine potential pilots.
- Interaction with clinicians and researchers to gain feedback from the pilots.
- Interaction with all relevant parties concerning legal and ethical issues identified by the pilots.
- Dissemination to the wider biomedical community on the outcome of the pilots.
- Interaction and exchange of results with similar groups worldwide.
- The formulation and specification of potential new applications in conjunction with the end user communities.

Recently, the Healthgrid association held its third annual international conference [36] at which the progress made in the spectrum of biomedical Grid projects was reviewed. The MammoGrid project provided important input to the ongoing debate on the role of Grids for (bio-)medical informatics. One very clear conclusion of the conference is the need to have greater involvement of the clinician community in the active use of medical informatics applications as demonstrated by MammoGrid.

Finally, Grid computing is a promising distributed computing paradigm that can facilitate the management of federated medical images. This technology spans locations, organizations, architectures and has the potential to provide computing power, collaboration and information access to everyone connected to the Grid. Grid-based applications like the MammoGrid project benefit from this solution being based on open-internet standards. These applications are potentially cross platform compatible, cross programming interoperable and widely accepted, deployed, and adopted.

**Acknowledgements**

The authors wish to thank their institute and the European Commission for support and to acknowledge the contribution of the following MammoGrid project members: Professor Roberto Amendolia (formerly of CERN), David Manset (at UWE for MammoGrid, now Maat GKnowledge), Dr Ruth Warren and Iqbal Warsi (Addenbrookes Hospital, Cambridge), Dr Chiara Del Frate and Professor Massimo Bazzocchi (Policlinico Universitario, Udine), Professor Sir Mike Brady and Chris Tromans (Oxford University), Martin Cordell (Mirada Solutions), Dr Piernicola Oliva (Sassari University), Drs Evelina Fantacci and Allessandra Retico (Pisa University) and Dr Jose Galvez and Dr Predrag Buncic (CERN). Furthermore thanks are extended to Jan Talmon and the Editorial Board for constructive comments in the preparation of this manuscript.